# Causal machine learning reveals age-dependent radiation dose effects on mandibular osteoradionecrosis

**Authors:** Jingyuan Chen, PhD[1], Yunze Yang, PhD[2], Olivia M. Muller, MD[3], Lei Zeng, MD[1], Zhengliang Liu, MS[4], Tianming Liu, PhD[4], Robert L, Foote, MD[4], Daniel J, Ma, MD[4], Samir H, Patel, MD[1], Zhong Liu, PhD[6,*] and Wei Liu, PhD[1,*]

[1]Department of Radiation Oncology, Mayo Clinic, Phoenix, AZ 85054, USA

[2]Department of Radiation Oncology, the University of Miami, FL 33136, USA

[3]Department of Dental Specialties, Mayo Clinic Rochester, Rochester, MN, 55905, USA

[4]Department of Computer Science, University of Georgia, Athens, GA, 30602, USA

[5], Mayo Clinic, Rochester, MN, 55905, USA

[6]Institute of Western China Economic Research, Southwestern University of Finance and Economics, Chengdu, Sichuan 611130, China

Co-Corresponding author: Wei Liu, PhD, Professor of Radiation Oncology, Department of Radiation Oncology, Mayo Clinic Arizona; e-mail: Liu.Wei@mayo.edu.

## Conflicts of Interest Disclosure Statement

No

## Funding Statement

This research was supported by NIH/NIBIB R01EB293388, by NIH/NCI R01CA280134, by the Eric & Wendy Schmidt Fund for AI Research & Innovation, and by the Kemper Marley Foundation.

## Ethical Approval

This study was approved by Mayo Clinic Arizona institutional review board (IRB#: 24-011106).

## Data Availability Statement

The data analyzed during the current study are not publicly available due to patient privacy concerns and institutional policies regarding protected health information (PHI). However, de-identified data that support the findings of this study are available from the corresponding author upon reasonable request and with appropriate institutional review board (IRB) approval.

## Acknowledgments

This research was supported by NIH/BIBIB R01EB293388, by NIH/NCI R01CA280134, by the Eric & Wendy Schmidt Fund for AI Research & Innovation, and by the Kemper Marley Foundation.

## Abstract

Distinguishing causal relationships from statistical correlations remains a fundamental challenge in clinical research, limiting the translation of observational findings into interventional treatment guidelines. Here we apply causal machine learning to establish causal effects of radiation dose parameters on mandibular osteoradionecrosis (ORN) in 931 head and neck cancer patients treated with volumetric-modulated arc therapy. Using generalized random forests, we demonstrate that all examined dosimetric factors exhibit significant positive causal effects on ORN development (average treatment effects: 0.092–0.141). Integration with explainable machine learning reveals substantial treatment effect heterogeneity, with patients aged 50–60 years showing the strongest causal dose–response relationships (conditional average treatment effects up to 0.229), while patients over 70 years demonstrate minimal effects. These results suggest that age-stratified treatment optimization and personalized treatment planning for the dosimetric factors could reduce ORN risk. Our findings demonstrate that causal inference methods can transform clinical retrospective radiotherapy data into personalized treatment recommendations, providing a methodological framework applicable to toxicity prediction across oncology and other clinical domains where treatment decisions depend on complex dose–response relationships.

## Introduction

The distinction between correlation and causation represents one of the most fundamental challenges in clinical patient outcome research[1,2]. While randomized controlled trials remain the gold standard for establishing causality, they are often impractical, unethical, or prohibitively expensive for many clinical questions - particularly those involving treatment toxicities that manifest years after intervention[3,4]. Consequently, clinical decision-making frequently relies on observational studies that identify statistical associations but cannot determine whether modifying a factor would actually change patient outcomes. This limitation has profound implications: interventions designed around correlational evidence may prove ineffective, wasteful, or even harmful[1,2,5].

Recent advances in causal machine learning offer a promising solution to this long-standing problem[6-13]. Grounded in the potential outcome framework and counterfactual reasoning[6-9], these methods enable estimation of causal treatment effects from observational data by explicitly modeling and adjusting for confounding variables. Generalized random forests (GRF) [10,12-14] provide a particularly flexible approach, accommodating continuous treatments, handling high-dimensional confounders without parametric assumptions, and estimating heterogeneous treatment effects across patient subgroups. Despite their transformative potential, these methods remain largely unexplored in clinical outcome research, where correlational analyses and predictive models continue to predominate

Radiation oncology exemplifies a field where causal inference could yield substantial clinical impact. Approximately half of all cancer patients receive radiotherapy during their disease course[15,16], and treatment planning fundamentally depends on understanding dose–response relationships for both tumor control and normal tissue toxicity. Yet nearly all radiotherapy outcome studies to date have been observational, yielding correlational findings that cannot definitively guide dose optimization. Consider a simplified illustration (Fig. 1a): observational data consistently show that tooth discoloration correlates with lung cancer incidence, but this association is non-causal, both are downstream effects of smoking, and dental interventions would not reduce cancer risk. Analogously, demonstrating that a dosimetric parameter correlates with toxicity does not establish whether reducing that parameter would decrease complication rates.

This challenge is particularly acute in head and neck (H&N) cancer [17,18] radiotherapy, where critical organs at risk (OARs), including the mandible, salivary glands, and pharyngeal constrictors, are inevitably exposed to significant radiation doses due to their anatomical proximity to tumor targets[19-22]. H&N cancer is a major global health burden, ranking among the most common malignancies worldwide, with an estimated approximately 660,000 new cases and 325,000 deaths annually[17,18]. Radiotherapy, particularly volumetric-modulated arc therapy (VMAT), is a cornerstone of H&N cancer management due to its high conformality and normal tissue sparing[15,16]. Despite these technological advances, treatment-related adverse events (AEs) in H&N cancer remain substantial because tumors frequently abut - and may infiltrate - critical OARs such as the mandible, salivary glands, larynx, pharyngeal constrictors, esophagus, and muscles of masticatory[19-22].

Mandibular osteoradionecrosis (ORN) represents one of the most debilitating late toxicities[23,24], involving progressive bone injury, impaired wound healing, and potential pathological fracture that substantially diminishes quality of life. Patient outcome study in radiotherapy is crucial to identifying risk factors, optimizing treatment protocols, reduce AEs, and improving patient quality of life[20,23-31]. Several clinical factors (such as, preexisting dental disease, chemotherapy, smoking, diabetes) have been associated with ORN incidence[32,33]. Extensive research has identified statistical associations between dosimetric parameters (such as V50Gy, V60Gy, and mean dose) and ORN incidence[32-38]. Several clinical factors (such as, preexisting dental disease, chemotherapy, smoking, diabetes) have also been associated with ORN incidence[32,33]. However, the causal effects of these factors, the degree to which dose reduction would actually decrease ORN risk, remain unquantified.

Here we integrate causal inference with explainable machine learning to address this gap. This retrospective study analyzed the cohort of 931 H&N cancer patients treated with VMAT at an anonymized institution from 2013 to 2019. The dataset includes mandibular ORN outcomes along with detailed dosimetric and clinical factors. We pursue three objectives: (i) to estimate causal effects of dosimetric factors on ORN risk, (ii) to identify patient subgroups with heterogeneous treatment responses, and (iii) to demonstrate the broader applicability of this causal inference framework for personalized radiotherapy optimization. By quantifying causal effects and elucidating key risk drivers, our findings aim to inform risk-adapted planning that reduces ORN while maintaining tumor control, ultimately improving patient quality of life., To our knowledge, this work represents the first application of causal machine learning to late radiation toxicity, establishing a methodological template transferable to other clinical domains where treatment decisions depend on complex, individualized dose–response relationships.

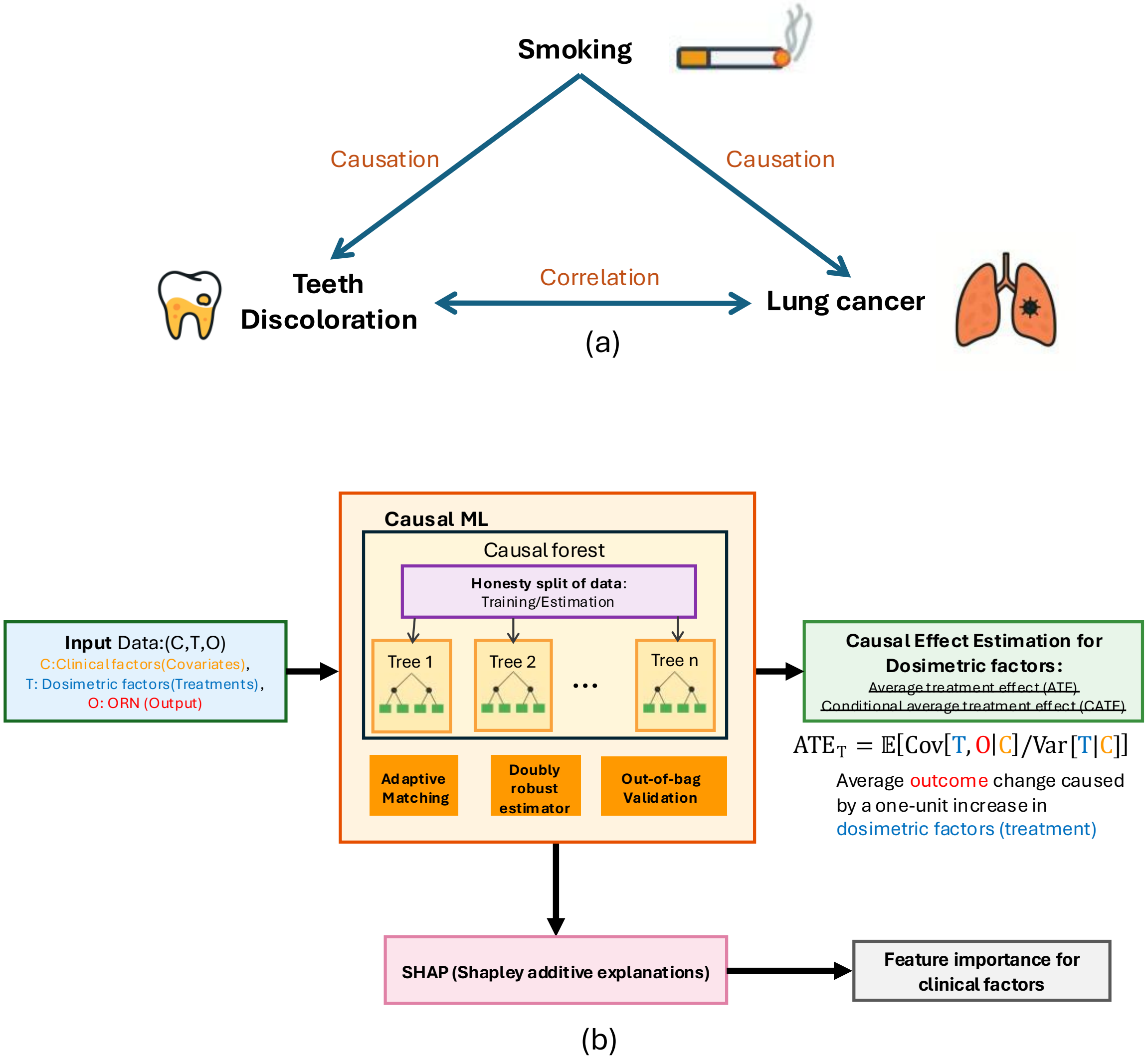

**Figure.1** (a) Illustration of the difference between *causation* and *correlation*. While yellow teeth and lung cancer exhibit statistical correlation, this association is non-causal and we cannot give interventional suggestions based on this correlation. Smoking acts as a common cause (confounding variable) that directly causes both yellow teeth and lung cancer. (b) The sketch of our approach for the causal analysis of the dosimetric factors and the importance analysis of the clinical factors. We employ the Causal Forest method, which is a tree-based causal machine learning method to reduce the bias from the real-word data and enable the estimation of the causal effects. The SHAP value and the importance based on the tree splitting frequency are adopted for the importance analysis of the clinical factors.

## Methods and Materials

### Patient Cohort

We included all consecutively treated, newly diagnosed head and neck (H&N) cancer patients at anonymized institutions A and B between April 2013 and August 2019 who received VMAT and had >2 years of follow-up, regardless of sex, age, minority status, vulnerable-population status, or body weight, and with a confirmed histologic diagnosis. The analytic dataset was restricted to patients who met all of the following criteria: (1) conventional fractionation of 120–220 cGy per fraction; (2) prescription dose ≥60 Gy to the high-risk target; and (3) for re-irradiated patients, negligible mandible dose on either the original or the re-treatment plan or documented ORN prior to re-irradiation. In total, 931 cases met these criteria. Demographic variables (e.g., age, sex) and relevant clinical information (e.g., prescription dose, tumor stage, concurrent chemotherapy, hypertension, diabetes, dental extraction, and smoking status [never, former, current]) were extracted (Supplemental Information [SI] Table 1). The study protocol was approved by the institutional review board (IRB). All patients granted permission to use their records for research purposes. The detailed information about diagnosis of ORN, treatment plans and dose calculation approaches are presented in the Supplementary Materials.

### Causal Inference

In causal-effect analyses, patient-level variables are commonly grouped into three categories. **Treatment variables** capture the exposure of interest, in this study, the dosimetric factors. **Covariates** are factors that may influence treatment assignment, outcomes, or both; here, these correspond to clinical factors. The **outcome** is the incidence of ORN. Our objective is to estimate treatment effects from observational data. We use the **Average Treatment Effect (ATE)** to quantify the population-level difference in potential outcomes under treatment versus no treatment.[8]

$$ATE = \mathbb{E}[O(T=1) - O(T=0)]$$

where $O(T=1)$ and $O(T=0)$ denote the potential outcomes, $O$, for the entire population under the treatment $(T=1)$ and control conditions $(T=0)$, respectively, $T$ represents the treatment assignment. When the outcome is normalized and the treatment is binary, the ATE corresponds to the change in the probability of the outcome occurring when the treatment is applied versus when it is not applied.

**Conditional average treatment effect for continuous variables**

In cases where the treatment is continuous, such as dose, ATE represents the average outcome change caused by a one-unit increase in treatment variables:[10,14]

$$ATE = \mathbb{E}\left[\frac{Cov[T, O|C]}{Cov[T|C]}\right]$$

where $Cov[\cdot,\cdot]$ refers to covariance operator and $C$ represents the covariates, which are the clinical factors. However, ATE only represents the populational causal effect. The Conditional Average Treatment Effect (CATE) extends beyond the population-level ATE by recognizing that treatments often affect different groups in distinct ways. [10,14]

$$CATE = \mathbb{E}\left[\frac{Cov[T,Y|C=c]}{Cov[T|C=c]}\right]$$

where $C=c$ represents conditioning on the specific subgroup characterized by clinical factors $C$ taking value $c$. While the ATE provides a single summary measure of a treatment's impact across an entire population, CATE captures the heterogeneity in treatment responses by estimating the average effect within specific subgroups. By conditioning on covariates like age and tumor stage, CATE transcends one-size-fits-all conclusions to reveal specifically who benefits most from the intervention.

**General random forest and robustness check**

Our analysis employed causal forest (CF) methodology to quantify treatment effects, particularly focusing on ATE estimation[10,12-14]. This approach adapts ensemble-based random forest techniques for causal analysis through the general random forest (GRF) framework, facilitating the identification of heterogeneous treatment responses across patient subgroups. The GRF algorithm constructs multiple decision trees; each trained on distinct data subsamples drawn without replacement. The splitting criterion at each node evaluates all possible partition points to optimize the separation of observations based on differential treatment responses. Notably, the algorithm prioritizes covariates that simultaneously influence both treatment assignment and outcomes, thereby addressing potential confounding effects from clinical factors. Out-of-bag validation was implemented to derive unbiased estimates of treatment effects and their standard errors, eliminating the need for separate validation datasets. This approach provided robust causal effect estimates for dosimetric parameters on ORN development while accounting for clinical variables. The details about robustness check are presented in the Supplementary Materials. Implementation was performed using the GRF package (version 2.4.0) in R statistical software (version 4.4.1).

**Clinical factor Importance**

We adopt two approaches to characterize the importance of the clinical factors: Shapley additive explanations (SHAP) importance and the GRF split-based importance. We integrated GRF into SHAP algorithm to calculate feature importance for clinical factors[39]. SHAP is a unified approach to explain the output of machine learning models based on game theory principles, specifically Shapley values. These values **quantify** each factor's contribution to the change in the expected model output when conditioning on that factor. The method is model-agnostic and maintains a strong theoretical foundation ensuring consistency. The SHAP value characterizes the correlation between the clinical factors and the patient outcomes.

The GRF split-based importance is defined as the weighted sum of how many times the clinical factor $C$ was split on at each depth in the forest[13].

$$I_{GRF}(C) = \frac{S_{i=c}}{\sum_i S_i}$$

Where $S_i$ represents the split times of the $i$th clinical factor and $S_{i=c}$ represents the split times of the specific clinical factor $c$. This importance measure captures the contribution of each clinical factor to the GRF model by tracking how frequently the factor is used to partition the data across all trees, with splits at shallower depths receiving higher weights to reflect their greater impact on the model. GRF split-based importance characterizes the contribution of the clinical factors to the heterogeneity of the treatment effects.

Following the global importance analysis, to investigate how individual clinical factors modulate treatment effects, we generated SHAP dependence plots. The SHAP dependence plots visualize how the model's predictions change across the range of factor values [40-43]. These plots reveal non-linear patterns and threshold effects that might not be apparent from population-level correlation analyses.

By combining GRF with SHAP values in our study, we were able to not only identify causal relationships between dosimetric factors but also quantify risky importance of clinical factor[44].

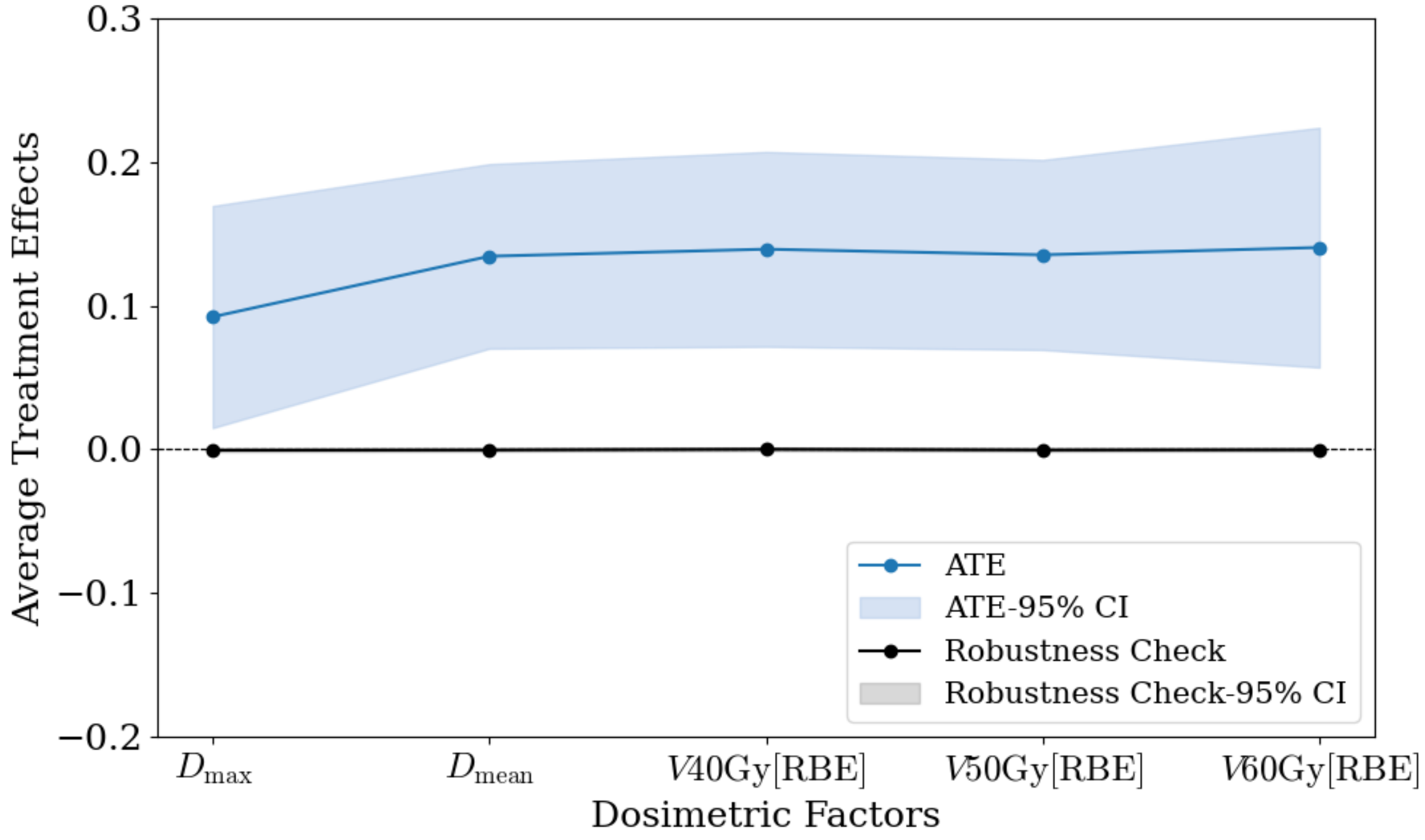

**Figure 2.** Average treatment effects of dosimetric factors obtained through the GRF method, with the method robustness check and 95% confidence intervals. The robustness check was calculated by replacing the actual data of each dosimetric factor with random numbers. Results closer to zero indicate greater robustness of the method.

## Results

### Average Treatment Effects of Dosimetric Factors

Figure 2 illustrates the ATEs of dosimetric factors obtained through the GRF method, with accompanying robustness validation. The analysis evaluated five key dosimetric parameters: maximum dose (Dmax), mean dose (Dmean), and three DVH indexes (V40Gy, V50Gy, and V60Gy). All dosimetric factors have positive ATE. Dmean, V40Gy, V50Gy, and V60Gy have a similar ATE. Dmean: 0.1344(95% CI: 0.0703–0.1986); V40Gy: 0.1393(95% CI: 0.0716-0.2071); V50Gy: 0.1354(95% CI: 0.0693-0.2015); V60Gy: 0.1405(95% CI: 0.0571-0.2239). The Dmax has the minimum ATE: 0.0922 (95% CI: 0.0150-0.1694).

The robustness check, displayed as the black horizontal line with gray confidence bands near zero, validated the reliability of our causal estimates. The absolute values of robustness check results are less than 0.001. When the actual dosimetric data were replaced with random numbers, the resulting effects converged to zero with narrow 95% confidence intervals, confirming that the observed treatment effects represent genuine causal relationships rather than methodological artifacts or chance findings. The clear separation between the actual ATEs and the robustness check results strengthens our confidence in the identified causal relationship.

These findings suggest that dosimetric factors, particularly Dmean, V40Gy, V50Gy, and V60Gy, have meaningful causal impacts on treatment outcomes.

### Clinical Factor Importance and SHAP Analysis

The importances or the correlation between the clinical factors and ORN were quantified by two approaches: 1) SHAP values and 2) the split-based importance from the GRF model. The results are shown in Figure 3.

Both important scores agree that age is the most important clinical factor, with a GRF split-based importance of 44% and SHAP value of 39%. Diabetes was identified as the least correlated to ORN, with a GRF split-based importance of 2% and SHAP value of 2%. These two approaches also show agreements on tumor stage, smoking history, and chemotherapy.

Overall, SHAP value and GRF split-based importance show similar values for five of the eight clinical factors all of which are lower than age. These results underscore the critical need for further analysis about age.

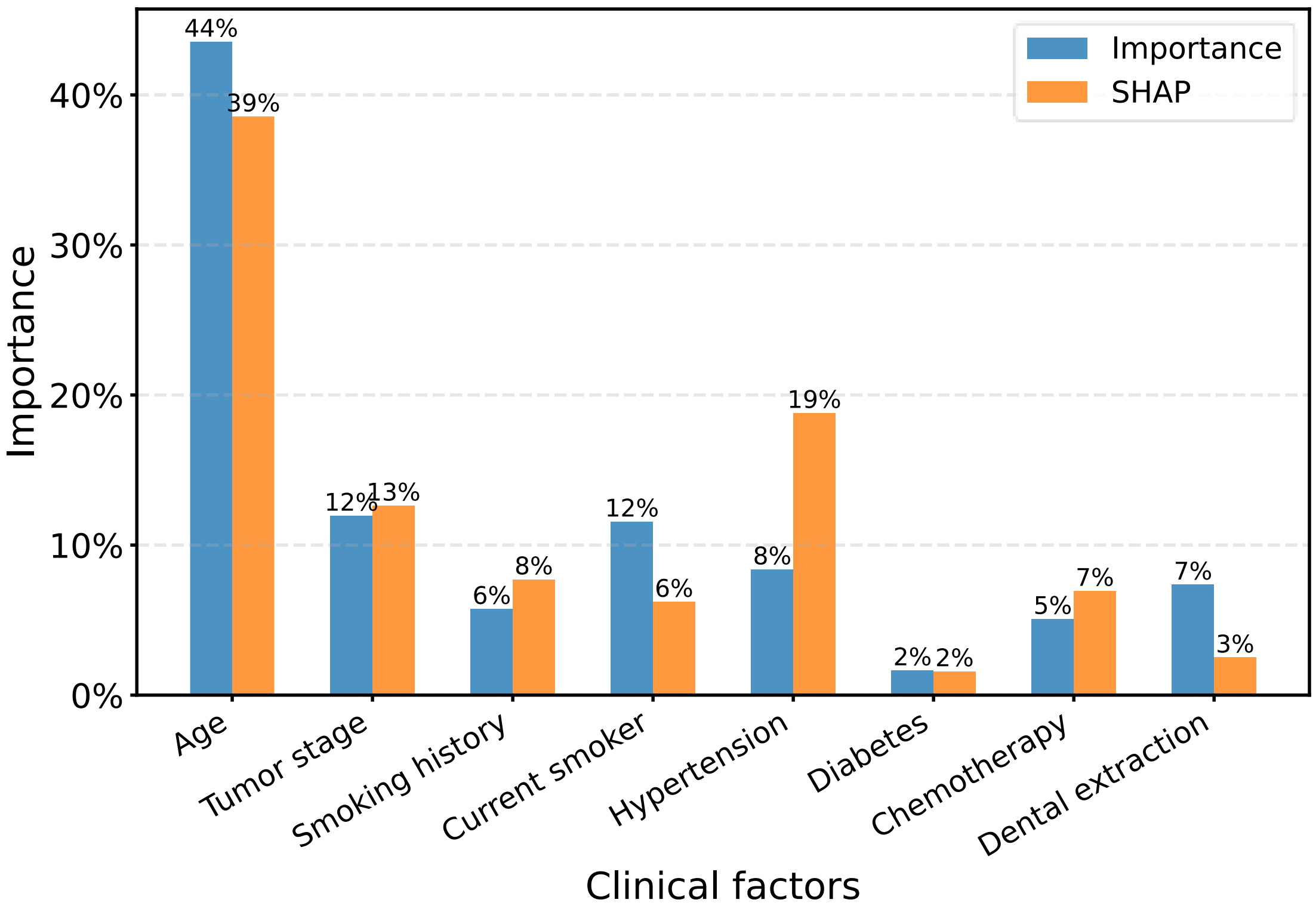

Figure 3. Importance of the clinical factors: The mean absolute SHAP value and the GRF split-based importance of the clinical factors. The SHAP value is obtained from the mean absolute values. The GRF split-based importance is based on how many times each clinical factor is split at each depth in the forest.

**Age-Dependent SHAP Value Analysis**

Figure 4 illustrates the SHAP value curve vs. age, revealing a distinctive non-linear relationship between patient age and treatment outcomes. The analysis demonstrates how age influences model predictions across the entire age spectrum, with individual patient data points (blue dots) showing the distribution of SHAP values at each age, and box plots summarizing the patterns for key age groups.

The SHAP values exhibit a clear inverted U-shaped relationship with age, characterized by three distinct phases. In the younger cohort (under 40 years), SHAP values remain lower value. A marked transition occurs in the 40-50 age group, where SHAP values increase substantially. The box plot for this cohort shows the highest SHAP value compared to other age subgroups, indicating the most significant correlation to the ORN incidence. Beyond age 60, a progressive decline in SHAP values is observed. The over-70 age group shows the lowest SHAP values. We should note that both SHAP value and GRF important score could only be interpreted as correlation instead of causation.

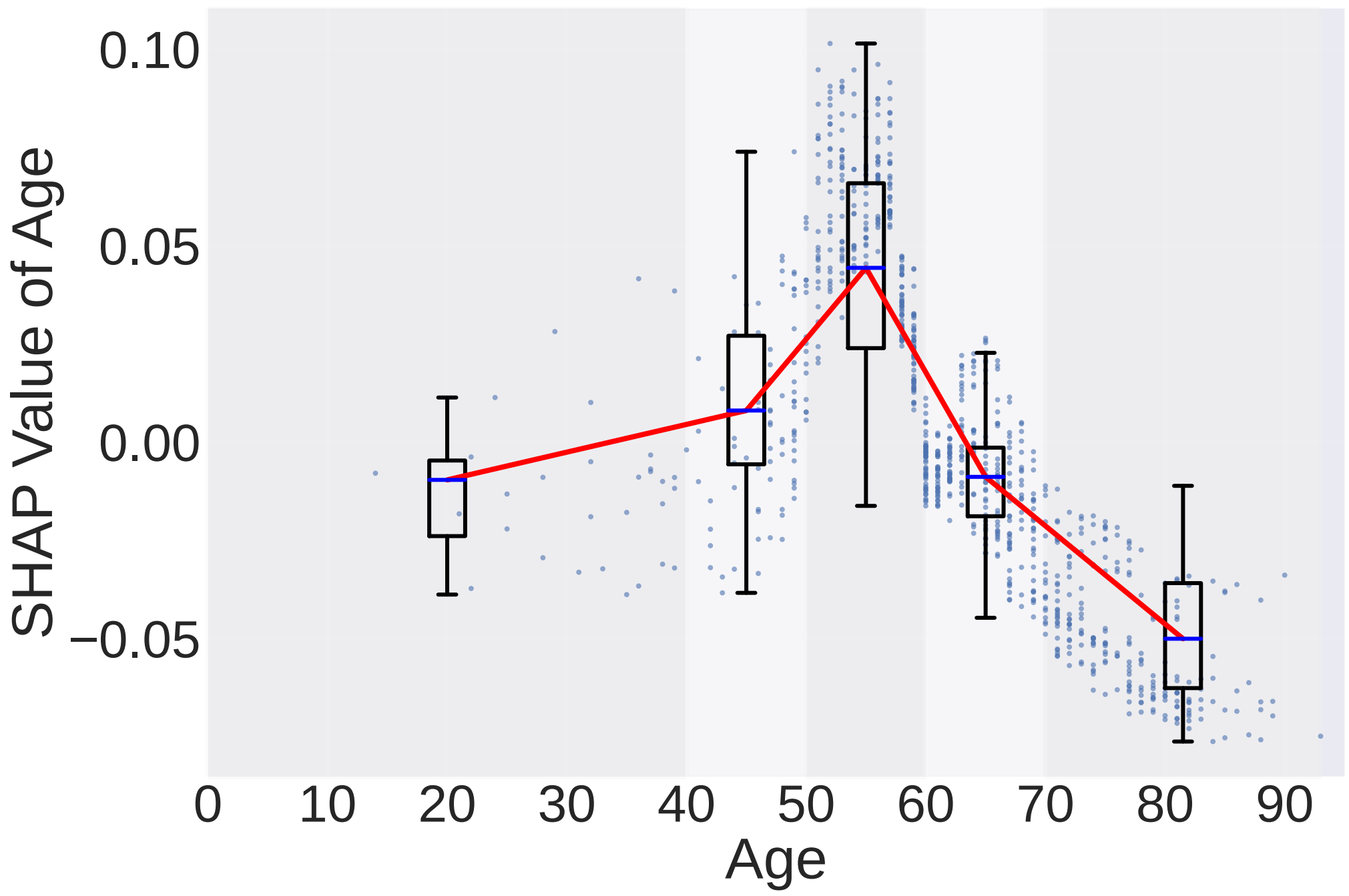

Figure 4. The SHAP value curve vs. Age. The blue points indicate SHAP value of each patient. The box plots are shown for the SHAP values of age in different age groups of under 40, 40-50, 50-60, 60-70, and over 70. Each box plot shows the interquartile range (IQR) with the median indicated by a blue line, whiskers extending to data points within 1.5×IQR. The red line connects the median SHAP values across all boxes, revealing the trend of feature importance as the feature value changes.

**Conditional Average Treatment Effects by Age Groups**

Based on the significant importance of age, we calculated the CATE of different dosimetric factor in different age subgroups. Figure 5 presents the CATEs of different dosimetric factors stratified by age groups, revealing substantial heterogeneity in treatment effects across different age cohorts.

The analysis examines dosimetric indices of Dmax, Dmean, V40Gy, V50Gy, and V60Gy) across four age groups: under 50, 50-60, 60-70, and over 70 years. A striking age-dependent pattern emerges across all dosimetric factors. The 50-60 age group consistently demonstrates the strongest treatment effects, with CATEs of 0.1803 (95% CI: -0.0481, 0.4087) for Dmax to 0.2289 (95% CI: -0.0013, 0.3993) for V50Gy. The concentration of high CATE values in this age range corroborates our earlier SHAP value findings, confirming that patients in 50-60 age cohert derive the greatest benefit from reducing the mandible dose in the VMAT treatment.

The youngest cohort (under 50 years) and the 60-70 years cohort exhibite moderate treatment effects across most dosimetric factors. The elderly cohort (70+ years) shows the most attenuated treatment effects, with CATEs ranging from 0.0242 for Dmax (95% CI: -0.0273, 0.0757) to 0.0832 for V60Gy (95% CI: -0.0321, 0.1986). Across all subgroups, V60Gy has the largest treatment effect and Dmax has the lowest treatment effect, which agrees with the ATE result.

The confidence intervals for several estimates have negative values. The 95% CI and negative CATE of Dmax for the under 40 age group could be optimized by a larger cohort with more events. This reflects a limitation of causal ML approaches, which require larger volume data than conventional statistical analysis.

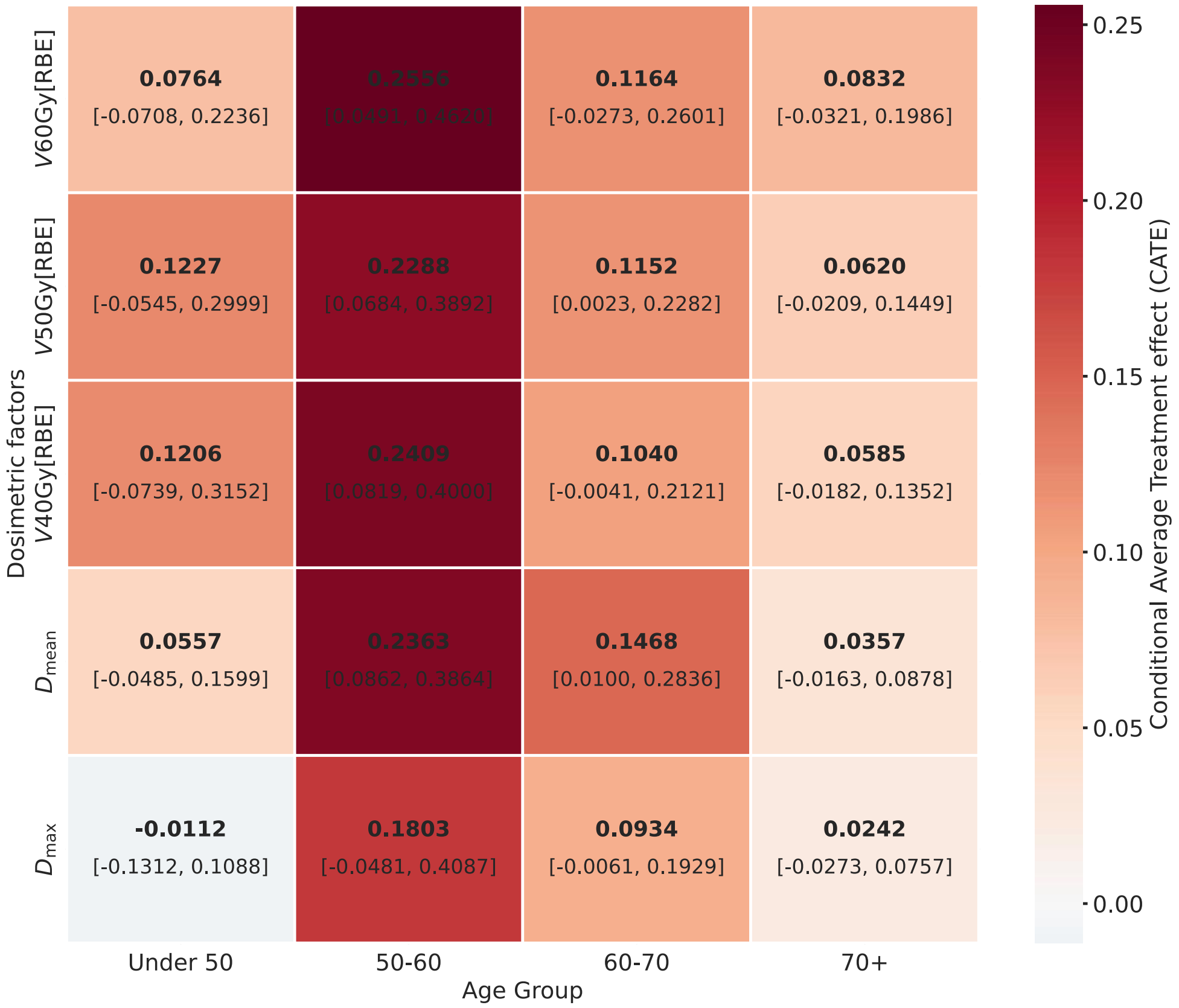

**Figure 5.** The conditional average treatment effect and 95% confidence intervals of the dosimetric factors $D_{max}$, $D_{mean}$, V40Gy, V50Gy and V60Gy by different age group. The ages are grouped by under 50, 50-60, 60-70 and over 70 years.

## Discussion

To the best of our knowledge, this causal-based analysis represents the first combination of explainable machine learning and causal machine learning to quantitatively interpret the dosimetric and clinical factors in radiotherapy patient outcome study. Through the GRF method, we have quantified the causal relationship between dosimetric factors and ORN. We also identified age as the most important clinical factor associated with ORN. Through CATE and SHAP value analysis, our results suggest that for patients aged 50-60 years old, dosimetric indices demonstrate the most significant causal relationship with ORN development. Therefore, for this age group, optimizing mandibular dose during radiotherapy treatment planning warrant particular attention. This study demonstrates the feasibility and value of causal machine learning for clinical outcome research, establishing a framework that advances beyond traditional correlational analyses toward actionable, personalized treatment recommendations.

We included 931 H&N cancer patients treated at our institution between 2013 and 2019 and the incidence rate of mandible ORN is found to be 5%. The important dosimetric factors and known clinical factors possibly related to mandible ORN are included in the study. Based on this large and comprehensive retrospective dataset and the novel causal ML methodology, we successfully quantify the causal relationship between dosimetric factors and ORN. All dosimetric factors showed positive ATEs, confirming their causal relationship with ORN. V40Gy, V50Gy, V60Gy and Dmean shows the comparable ATE, while Dmax shows less causal relationship with ORN. This suggests that the mean dose and the intermediate doses of mandible are more important than the maximum dose in respect to minimizing the incidence rates of ORN, which agrees with the reported results[25,34,45-49].

Another important innovation of our work is to demonstrate the feasibility of using causal methods to provide evidence for personalized treatment plans in radiotherapy. We creatively integrate causal machine learning with explainable machine learning to produce importance rankings for clinical factors. We stratified the most important clinical factor, age, and then further computed the CATE of dosimetric factors of different age groups for ORN. Our results show that, within the 50–60 age group, various dosimetric factors have a more pronounced causal effect on the occurrence of ORN; this is consistent with and explains the SHAP value curve vs. age derived from the explainable machine learning.

A key point of interpretation is the distinction between importance, causation, and association for clinical factors. Clinical factor importance related to SHAP values derived from explainable machine leaning methods should be understood as evidence of a strong statistical association between age and ORN—not as a proof that age *causes* ORN. Our subsequent CATE analysis further clarifies this distinction: the strong association observed for the 50–60-year subgroup with ORN arises because, within this age range, dosimetric factors have stronger *causal* effects on ORN. This underscores the necessity of employing causal machine-learning approaches to quantify causal effects rather than relying on association-based SHAP value alone derived from explainable machine leaning methods.

**Limitations**

Despite the valuable insights gained, several limitations warrant consideration. First, causal machine learning approaches typically require substantially larger data volumes, and our relatively small sample size resulted in wider 95% confidence intervals. Second, while we endeavored to satisfy the three key assumptions of causal inference (independence, ignitability, and positivity), achieving complete fulfillment would require even more comprehensive clinical databases. Given that ORN is a late-occurring complication, study with longer follow-up might provide more definitive results. Finally, while our approach represents methodological advancement, prospective validation in independent patient cohorts would further strengthen our findings.

## Conclusion

This study demonstrates that causal machine learning can bridge the gap between observational data and actionable clinical recommendations—a fundamental challenge that has long limited the translation of correlational findings into personalized treatment strategies. By pioneering the application of causal inference to radiotherapy toxicity, we establish causal relationships between dosimetric factors and mandibular ORN in 931 head and neck cancer patients treated with VMAT, advancing beyond traditional correlative analyses to provide actionable evidence for treatment optimization. Key findings include significant causal effects of dosimetric indices on ORN development (ATE: 0.09-0.14), with marked heterogeneity across age groups. Patients aged 50-60 years demonstrated the strongest treatment effects (CATE >0.22 for dosimetric indices), while those over 70 showed minimal response, indicating the need to further explore the concept of using age to further personalize treatment planning in radiotherapy.